\documentclass[pra,twocolumn,english,superscriptaddress,floatfix]{revtex4}
\usepackage{graphicx}
\usepackage{bm, amsmath, amssymb}
\usepackage{color}
 \usepackage{soul}
\usepackage{pdfpages}
\usepackage[T1]{fontenc}
\usepackage[latin9]{inputenc}
\usepackage{amstext}
\usepackage{bbold}
\usepackage{esint}

\def\be{\begin{equation}}
\def\ee{\end{equation}}
\def\bea{\begin{eqnarray}}
\def\eea{\end{eqnarray}}

\usepackage{babel}

\begin{document}

\title{Quantum caustics in resonance fluorescence trajectories}

\author{M. Naghiloo}
\affiliation{Department of Physics, Washington University, St.\ Louis, Missouri 63130}

\author{D. Tan}
\affiliation{Department of Physics, Washington University, St.\ Louis, Missouri 63130}

%\affiliation{Quantum Nanoelectronics Laboratory, Department of Physics, University of California, Berkeley CA 94720}
\author{P. M. Harrington}
\affiliation{Department of Physics, Washington University, St.\ Louis, Missouri 63130}

\author{P. Lewalle}
\affiliation{Department of Physics and Astronomy, University of Rochester, Rochester, NY 14627}

\author{A. N. Jordan}
\affiliation{Department of Physics and Astronomy, University of Rochester, Rochester, NY 14627}
\affiliation{Center for Coherence and Quantum Optics, University of Rochester, Rochester, NY 14627}
\affiliation{Institute for Quantum Studies, Chapman University, Orange, CA 92866}

\author{K. W. Murch}
\affiliation{Department of Physics, Washington University, St.\ Louis, Missouri 63130}
\affiliation{Institute for Materials Science and Engineering, St.\ Louis, Missouri 63130}

\begin{abstract}
We employ phase-sensitive amplification to perform homodyne detection of the resonance fluorescence from a driven superconducting artificial atom. Entanglement between the emitter and its fluorescence allows us to track the individual quantum state trajectories of the emitter conditioned on the outcomes of the field measurements. We analyze the ensemble properties of these trajectories by considering trajectories that connect specific initial and final  states. By applying the stochastic path integral formalism, we calculate equations-of-motion for the most likely path between two quantum states and compare these predicted paths to experimental data.  Drawing on the mathematical similarity between the action formalism of the most likely quantum paths and ray optics we study the emergence of caustics in quantum trajectories---places where multiple extrema in the stochastic action occur.  We observe such multiple most likely paths in experimental data and find these paths to be  in reasonable quantitative agreement with theoretical calculations.
\end{abstract}

\maketitle

\section{Introduction}
In ray optics, light travels the shortest optical path between two points. While this minimizes the action associated with the path, multiple minima in the action may occur under some circumstances. These phenomena, known as caustics, are described by catastrophe theory \cite{berry1980iv,arno13singularities}, which deals with discontinuous and divergent phenomena and is applicable to topics ranging from biology to social science \cite{posto2014book}. Caustics are well-studied in optics and have been extended to stochastic media for which the conditions are described by fixed statistical models \cite{berry1980iv, whit84stochastic}. In analogy to the  trajectory of light propagating in a turbulent medium, a continuously-monitored quantum system also undergoes stochastic trajectories in its quantum state space due to the back-action of continuous measurement \cite{wisebook,hatr13,murc13traj}. Similary for rays in optics, stochastic path integral formalism has been applied to the dynamics of diffusive quantum trajectories, revealing optimal quantum paths connecting two points in quantum state space \cite{chan13,webe14,chan15}.
%Path integrals may be applied to a wide range of physical phenomena in quantum or statistical mechanics.

The probability density associated with a particular quantum trajectory may be described with a stochastic action; just as an action may be optimized to obtain classical paths in Lagrangian or Hamiltonian mechanics, the stochastic action in a path integral may be extremized to obtain optimal paths through the state space of a measured quantum system.
Such most-likely paths (MLPs) have been studied in the context of continuous quantum non-demolition and fluorescence measurements  \cite{chan13,webe14,jord15}, quantum entanglement \cite{chan16}, and the stochastic path integral formalism has been applied to calculate correlation functions in measurement observables \cite{chan15}.  More broadly, MLPs have been applied to general diffusion processes \cite{lang78} and the case of chemical kinetics where multiple MLPs have been predicted for classical stochastic dynamics \cite{dykm94,kame11}.

Here we report on the  observation of individual quantum trajectories via homodyne monitoring of resonance fluorescence from a superconducting qubit. By analyzing trajectories meeting the same pre- and post- selection, we identify optimal quantum paths which are in agreement with the paths predicted from the path integral formalism.  As with optical caustics, the path integral formalism predicts the existence of multiple optimal paths for quantum trajectories \cite{phil2016}. Under driving conditions where such {\it quantum caustics} are predicted to occur, we find that our observed quantum trajectories naturally split into two communities \cite{reic06,newm06,fort10}. We show number of captured trajectories in each community and associated MLPs are in quantitative agreement with the theory prediction.
Our results provide insight into the fundamental light matter interaction of resonance fluorescence and highlight the connection between quantum dynamics and divergent systems, offering routes to investigate the possibility of chaos and understand the quantum classical boundary.

%\section{Experiment setup }
%\section{Experimental system}
\section{Quantum trajectories in resonance fluorescence}
\begin{figure}[]
  \begin{center}
    \includegraphics[width=0.45\textwidth]{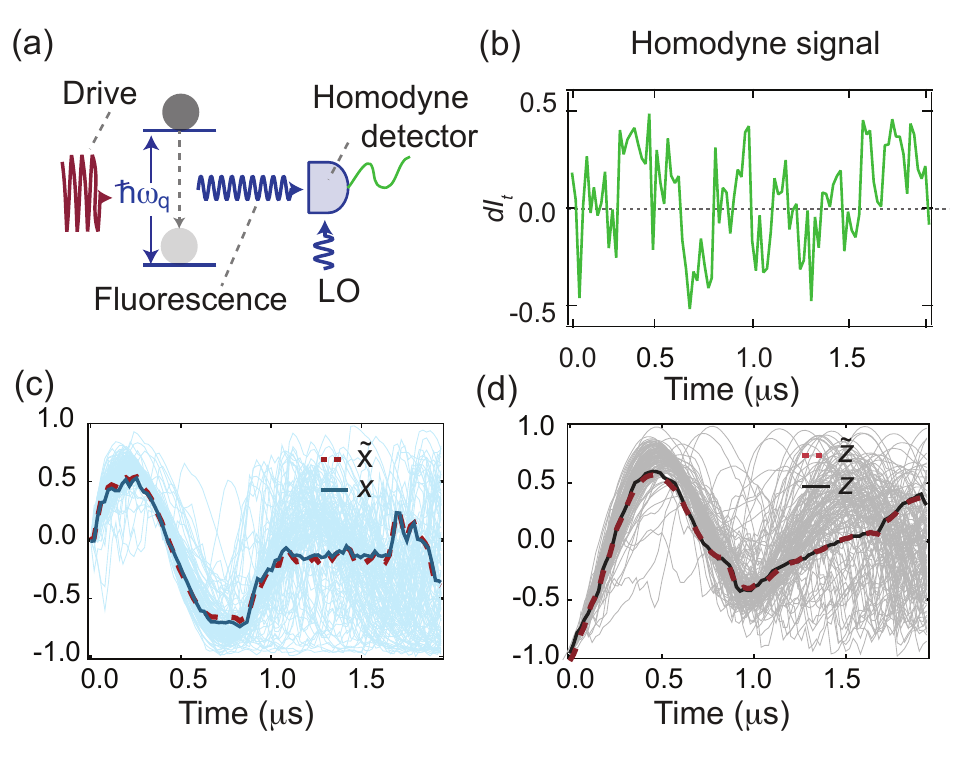}
  \end{center}
  \vspace{-.2in}
  \caption{\footnotesize \textbf{Resonance fluorescence quantum trajectories.} (a),~The experiment uses a near-quantum limited Josephson parametric amplifier to perform homodyne measurement of the fluorescence emitted by an effective two-level system which is realized by  resonant interaction of a transmon circuit and a 3D aluminum cavity. (b), The dimensionless homodyne signal (denoted $dI_t$ at time step $t$) reflects the quantum fluctuations of the measured electromagnetic mode and is normalized so that its variance is $\gamma\, dt$. (c,d),   The $x$ and $ z$ components of several trajectories calculated using the SME and homodyne signal. A specific trajectory ($\tilde{x},\tilde{z}$) is compared to its tomographic reconstruction ($x,z$). The close agreement between the curves indicates that the SME faithfully tracks the quantum state.}\label{fig1}
\end{figure}

Our experiment focuses on the stochastic trajectories of a driven quantum emitter that interacts radiatively with its environment. The combination of resonant drive and decay---resonance fluorescence---results in emitted light with uniquely quantum features that have been studied extensively in atomic and condensed matter systems \cite{schu74,kimb77,wall81,hoff97, koch95, camp14, Cars15, toyl16}. In departure from conventional studies of resonance fluorescence, here we approach the quantum dynamics of resonance fluorescence in the context of quantum measurement and use the fluorescence signal to calculate quantum trajectories for the emitter's state. The experiment consists of  an effective two-level system (qubit) resulting from the resonant interaction between a transmon circuit \cite{koch07} and a 3D aluminum cavity \cite{paik113D}. The qubit decay is set by deliberate coupling of the cavity to a 50 $\Omega$ microwave transmission line allowing the fluorescence signal to be collected with high efficiency   (Fig.\ \ref{fig1}a).  The effective Hamiltonian of the system under the rotating wave approximation is   ($\hbar=1$),   
\begin{equation}
H= - \frac{ \omega_q}{2}\sigma_z - \frac{\Omega}{2} \sigma_y + \gamma(a^{\dagger} \sigma_- + a \sigma_+), 
\label{ham}
\end{equation}
where the three terms in the Hamiltonian describe the qubit, the drive, and the resonant interaction of qubit and environment, respectively. Here, $a(a^{\dagger})$ is annihilation(creation) operator for the quantized field in the transmission line and the qubit is a pseudo-spin system described by the Pauli spin operators $\sigma_x,\sigma_y,\sigma_z$ and $\sigma_\pm$.

%\section{Quantum trajectory via resonance florescence}
We use a near-quantum-limited Josephson parametric amplifier to perform phase sensitive amplification of a single field quadrature $\propto (a^\dagger e^{i\phi} + a e^{-i \phi})$ of the fluorescence. By virtue of the interaction Hamiltonian, this quadrature of the field is correlated with a specific dipole of the emitter, $\sigma_-e^{i\phi} + \sigma_+ e^{-i \phi}$. Measurements of this quadrature amplitude therefore convey information about the emitter and can be used to reconstruct the evolution of the emitter's state \cite{nagh16}.   Under appropriate scaling, the digitized measurement signal is represented by (for $\phi=0$), %We adjust the phase of amplification and scale the signal so that the unit-less homodyne signal is a noisy estimate of $\langle x \rangle$ at each time step (Figure \ref{fig1}b) and represented by
\begin{equation} \label{dIt}
dI_t = \sqrt{\eta}\gamma \langle\sigma_x \rangle dt + \sqrt{\gamma} \xi_t\,dt.
\end{equation}
The measurement signal is proportional to $\langle \sigma_x \rangle$, but also contains a zero-mean random increment, $\xi_t$, with variance of $dt^{-1}$, arising from the quantum fluctuations of the field.  Here, $\eta = 0.45$ is the quantum efficiency and $\gamma=1.42$ $\mu$s$^{-1} $ is the decay rate of the emitter which characterizes the measurement strength \cite{nagh16}.  The random increments $\xi_t$ imply a  stochastic evolution for the emitter's state, represented by density matrix $\rho$, which can be calculated by the corresponding stochastic master equation (SME) \cite{gamm13},
\begin{equation} \label{SME1}
\dot{\rho} =  -i \frac{\Omega}{2} [\sigma_y,\rho]  +  \gamma \mathcal{D}[\sigma_-] \rho  + \sqrt{\gamma \eta } \mathcal{H} [\sigma_- ] \rho \xi_t ,
\end{equation}
where $\mathcal{D}[\sigma_-]$ and $\mathcal{H}[\sigma_-]$ are the dissipation and jump superopertators respectively \cite{gamm13}. We may recast this stochastic master equation  in Bloch vector components, $x\equiv\langle\sigma_x\rangle$ and  $z\equiv\langle\sigma_z\rangle$, as follows,
 \begin{equation} \label{SME2}
 \begin{split}
 \dot{z} =& +\Omega x  +  \gamma (1-z)  + \sqrt{\eta \gamma} x (1-z) \xi_t,  \\
 \dot{x} =& - \Omega z  - \frac{\gamma}{2} x  + \sqrt{\eta} ( 1-z - x^2  )\xi_t,  \\
\end{split}
\end{equation}
where the initial state, homodyne measurement phase, and drive Hamiltonian (characterized by a Rabi frequency $\Omega/2\pi = 0.9$ MHz) allow us to restrict the evolution to the $x$--$z$ plane of the Bloch sphere.  % Figure \ref{fig1}c and d show several trajectories for both $x$ and $z$ components of the qubit evolution.

Figure 1 depicts the quantum trajectories that are calculated for the emitter as it evolves due to both unitary drive and radiative decay. The quantum trajectories exhibit stochastic features associated with the back-action of homodyne measurement of the fluorescence \cite{nagh16} and evolution originating from the unitary drive. %The calculated trajectories are verified by quantum state tomography (Figure 1c,d) as evidenced by the close agreement between a specific trajectory and the average outcomes of projective measurements conditioned on the trajectory.
The calculated quantum trajectories are validated with quantum state tomography to verify the predicted expectation values $\langle \sigma_x \rangle$, $\langle \sigma_z\rangle$  are in agreement with the average outcomes of projective measurements. Figure 1c,d show the close agreement between a specific trajectory and reconstructed trajectory by the state tomography technique  described in previous work \cite{murc13traj,webe14,nagh16}.

\begin{figure}[]
  \begin{center}
    \includegraphics[width=0.45\textwidth]{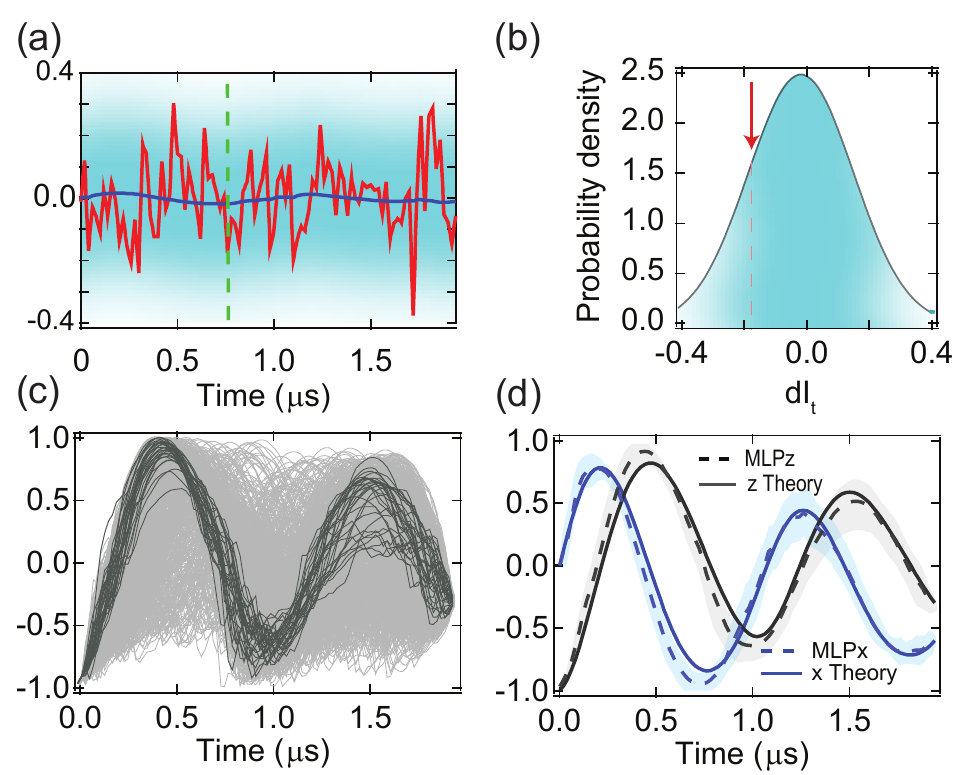}
  \end{center}
  \vspace{-.2in}
  \caption{\footnotesize \textbf{Path probability and most likely paths.} (a), Demonstration of path-probability calculation. The blue curve indicates $\sqrt{\eta} \gamma  x\, dt$ which determines the probability distribution (blue background color) for a corresponding homodyne signal (red curve). (b), A cross-section of the probability density at the time indicated by the dashed green line of (a). The Gaussian distribution is shifted from zero by $\sqrt{\eta} \gamma x\, dt$. The arrow shows the value of homodyne signal at that point. (c), The pre- and post-selected trajectories for initial $\{z(0),x(0)\}=\{-0.97,0\}$ and final $\{z( 1.94\ \mu\mathrm{s} ),x(1.94 \ \mu\mathrm{s})\}=\{-0.3,-0.6\}$ boundary condition (only $z$ traces are shown). The top five percent of which have lowest Euclidean distance from others are highlighted in black. (d), Solid black(blue) line is the theoretical MLP for $z (x)$ obtained from Eqs.\ (6). The dashed lines are the corresponding experimental MLP obtained by averaging the highlighted trajectories. The shaded area along experimental curves shows the standard deviation of the averaged trajectories.}\label{fig2}
\end{figure}
 
%\section{Most likely path}
\section{Most likely quantum path}

We return to the analogy between the stochastic trajectories of the emitter's state in quantum phase space and stochastic trajectories in optics such as starlight twinkling through a turbulent atmosphere. In both cases the randomness immediately suggests the question of statistical character. What is the most probable path for the system for a given initial and final state?  To address this question we examine the relation  between the homodyne signal  and the state coordinate $x$ (Eq.\ \ref{dIt}). As shown in Fig.\ \ref{fig2}, the detected signal is Gaussian distributed about $x$; signals with values in a region close to $\sqrt{\eta}\gamma x\, dt$ are more probable than signals in that  vary significantly from this mean value.  In a given trajectory, the measurement results lead to stochastic evolution of the state coordinates $(x,z)$, but the distribution of measurement results follows the evolution of $x$; the detection signals and state trajectories are thus coupled, leading to rich phenomena associated with state and signal correlations \cite{foro16,xu15,jord05,jord15}.

Considering this state-signal correlation, we are able to calculate the joint probability density for each measurement signal ($dI_t$) and state coordinates  $\mathbf{q}\equiv (x,z)$ with initial state $\mathbf{q}_0$ and final state $\mathbf{q}_N$,
\begin{eqnarray} \label{joint_prob}
P_\mathrm{joint}&=& \delta^{(2)}(\mathbf{q}-\mathbf{q}_0) \delta^{(2)}(\mathbf{q}-\mathbf{q}_N) \prod_{n=0}^{N-1} \mathcal{P}_n(dI_t|\mathbf{q}). 
\end{eqnarray}
Where $\mathcal{P}_n(dI_t|\mathbf{q})$ is the probability density for signal $dI_t$ at time $t = n\, dt$ and  $N$ is the total number of time steps. The most likely path maximizes the total path probability density $P_\mathrm{joint}$. To find this path we introduce a stochastic path integral representation of the joint probability, $P_\mathrm{joint} \propto \int \mathcal{D{\bf p}}  \ e^\mathcal{S}$. The term $\mathcal{S}$ is the stochastic action and $\mathcal{D{\bf p}} $ is an integral measure over the conjugate variables $(p_x ,p_z)$ which are auxiliary dynamical parameters that impose
the correct quantum measurement back-action dynamics \cite{chan13,chan15}. This approach has been verified experimentally for continuous quantum non-demolition measurement \cite{webe14}. %By adopting  the path integral formalism for homodyne detection of resonance fluorescence \cite{jord15}.\\%  we compute a ``stochastic Hamiltonian'' $H$; MLPs are solutions to the dynamical system generated by $H$ (see the SI for details).\\
Here we adopt the path integral formalism for homodyne detection of resonance fluorescence to obtain a corresponding stochastic Hamiltonian whose solutions (via Hamilton's equations) are the most-likely paths (MLPs). This is related to the stochastic action, $\mathcal{S}= \int [ -\dot{\textbf{q}} \cdot \textbf{p} +\mathcal{H}] dt$, as discussed at length in \cite{chan13,chan15,jord15,phil2016}. The full stochastic Hamiltonian for our driven-fluorescence system, using all three dimensions $x \in [-1,1]$, $y \in [-1,1]$ and $u \in [0,2]$ ($u=1+z$, for compactness) of the Bloch sphere, may be derived using the master equation (Eq.\ \ref{SME1}) and methods described in Ref. \cite{jord15}, which yield
\be\begin{split}
\mathcal{H} =& p_u \bigg[ \Omega x +u \gamma\left(1-2\eta + \frac{\eta u }{2} \right) \\&+\sqrt{\eta\gamma} x r (2-u) + 2 \gamma (\eta-1)\bigg] \\ &+ p_x \bigg[ -\Omega(1-u) - \frac{\gamma}{2} x (1+\eta u- 2\eta ) \\ &+ \sqrt{\eta \gamma} (2-u - x^2)r \bigg] \\ & + p_y \left[ - \frac{\gamma}{2} y (1+\eta u-2\eta) - \sqrt{\eta \gamma} x y r \right] \\ & - \frac{r^2}{2} + r \sqrt{\eta \gamma} x - \frac{\eta \gamma u }{2}.
\end{split}\ee
The stochastic readout is given by $r$, $\gamma$ is the fluorescence rate, $\eta$ the measurement efficiency, and $\Omega$ denotes the Rabi frequency.

\par For our case, where the dynamics is restricted to the $x$--$z$ plane of the Bloch sphere, we may eliminate $\dot{y}$ and $\dot{p}_y$ by setting $y = 0$ (setting $y=0$ decouples $p_y$ from the remaining equations regardless of its value). This simplification leads to a reduced stochastic Hamiltonian inhabiting a four, rather than six-dimensional phase space:
\be\begin{split}\label{h4}
H =& p_u \bigg[ \Omega x +u \gamma\left(1-2\eta + \frac{\eta u }{2} \right) \\&+\sqrt{\eta\gamma} x r (2-u) + 2 \gamma (\eta-1)\bigg] \\ &+ p_x \bigg[ -\Omega(1-u) - \frac{\gamma}{2} x (1+\eta u- 2\eta ) \\ &+ \sqrt{\eta \gamma} (2-u - x^2)r \bigg] \\ & - \frac{r^2}{2} + r \sqrt{\eta \gamma} x - \frac{\eta \gamma u }{2}.
\end{split}\ee
Since $H$ is time-independent, each MLP conserves a ``stochastic energy'' $E = H$. The deterministic equations-of-motion are %, the deterministic equations-of-motion for the MLP are \cite{jord15}
\begin{subequations}\label{eq-mlp2witheta}
\begin{eqnarray}
{\dot z} = +\Omega x + \gamma (1-z) \left(1 - \frac{\eta (1-z)}{2}\right) 
+ \sqrt{\eta \gamma} x (1-z) r, \label{udot}\quad \ \\
{\dot x} = -\Omega z-\frac{\gamma}{2} x (1 - \eta (1-z))
+\sqrt{\eta \gamma} (1-z - x^2) r, \label{xdot} \quad \quad \ \  \\
{\dot p}_z = -\gamma p_z (1 - \eta (1-z) + \sqrt{\eta \gamma} x r)  \quad \quad \quad \quad \quad \quad \quad \quad \quad 
 \nonumber \\
+p_x (\Omega + \gamma \eta x/2 + \sqrt{\gamma \eta} r )  - \eta \gamma/2, \quad \quad \quad  \\
{\dot p}_x  =   -p_z ( \Omega  + \sqrt{\eta \gamma} (1-z) r) + p_x ( \gamma(1 - \eta (1-z))/2  \quad \ \ \nonumber \\
+ 2 \sqrt{\eta \gamma} x r)  - \sqrt{\eta \gamma} r, \quad \quad \quad \quad
\end{eqnarray}
\end{subequations}
where $r = \sqrt{\eta \gamma} [x + p_x (1-z - x^2) + p_z x (1-z) ]$ is the (deterministic) optimal signal that replaces the homodyne signal (Eq.\ \ref{dIt}) $dI/dt$. The first two equations are comparable with the SME we use to update the experimental quantum trajectories (in Stratonovich form), and the last two equations pertain to the auxiliary parameters. The four equations-of-motion for the most likely path involving $z, x, p_z, p_x$ in Eqs.\ (\ref{eq-mlp2witheta}), combined with the constraint for $r$, may be solved numerically given initial and final states as we show in Fig.\ \ref{fig2}d (solid curves) for initial $\{z(0),x(0)\}=\{-0.97,0\}$ and final $\{z( 1.94\ \mu\mathrm{s} ),x(1.94 \ \mu\mathrm{s})\}=\{-0.3,-0.6\}$). Analytical solutions for these equations-of-motion for unity quantum efficiency and pure state evolution are discussed in Appendix \ref{ap_pure}.

 \begin{figure}[]
  \begin{center}
    \includegraphics[width=0.45\textwidth]{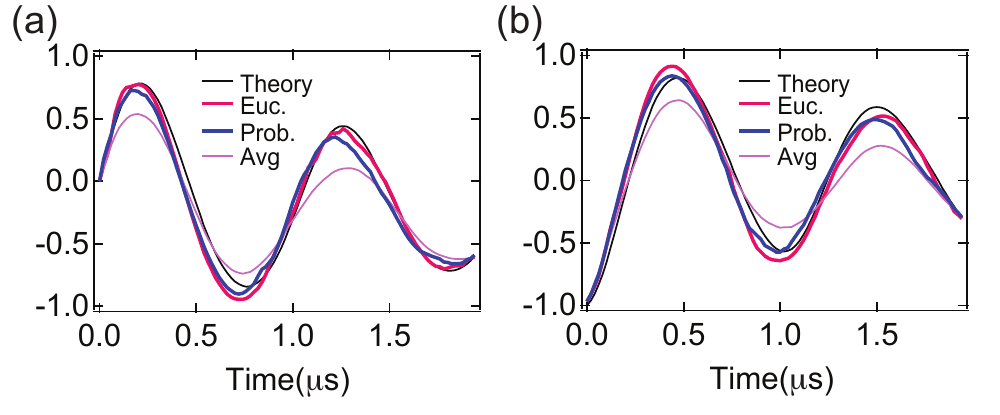}
  \end{center}
  \vspace{-.2in}
  \caption{\small  \textbf{Comparison of different experimental measures for most likely paths.} We compare the two different measures for experimental MLPs to theory for for initial $\{z(0),x(0)\}=\{-0.97,0\}$ and final $\{z(1.94\ \mu \mathrm{s}),x(1.94\ \mu \mathrm{s})\}=\{-0.3,-0.6\}$ states (panels (a,b) show trajectories for $x,z$ respectively).  The Euclidean distance method (red) and path probability density maximization (blue) are both in close agreement with theory (black).  The average of the full ensemble of pre- and post-selected trajectories (gray) deviates significantly from the MLP. }\label{fig6}
\end{figure}

\begin{figure*}[t]
  \begin{center}
    \includegraphics[width=0.95\textwidth]{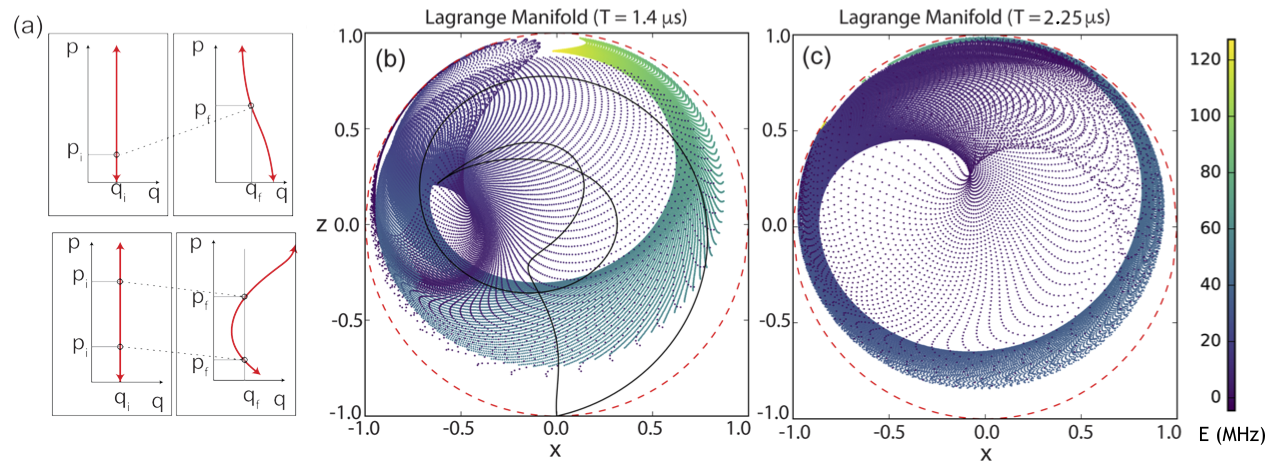}
  \end{center}
  \vspace{-.2in}
  \caption{\footnotesize  \textbf{The Lagrange Manifold.} (a), Schematic representation of Lagrange manifold folding leading to multiple most likely paths. For an initial state $\{z_i,x_i\}$ we consider all possible initial momenta, represented as a vertical red line. Evolution under Eqs.\ (\ref{eq-mlp2witheta}) results in different final states for different initial momenta specifying the most likely path that connects initial and final states, as shown in the left panels.  When the red curve fails the vertical line test as shown in the right panels, multiple initial momenta result in the same final state.  (b,c), The Lagrange manifold used to find MMLPs in this system is a two-dimensional object in a four-dimensional Hamiltonian phase space; we show its projection into the $x$--$z$ plane of the Bloch sphere. A sampled range of momenta $\{p_{zi},p_{xi}\}$ are initialized for the state ($\{z_i,x_i\} =\{-1,0\}$) at $t=0$ and allowed to evolve up to $T = 1.4$ $\mu s$ (b) and $T = 2.25$ $\mu s$ (c); each point in the scatter plot corresponds to a path generated by a different initial momentum. Compare with panel (a); just as a line of all $p$ may intersect the manifold one or more times at a given $q_f$, we may now imagine sticking a pin into the figure at a particular point $\{z_f,x_f\}$, which might go through one or more layers of the manifold. Each intersection or layer corresponds to a distinct path reaching the chosen final state. Caustic regions emenating from a catastrophe in the manifold are clearly visible at both of the times shown above. The counter-clockwise spiraling of the manifold is the result of the Rabi drive applied to the qubit; we see that overlap in the manifold due to this spiraling matures into a cusp catastrophe over time. Color denotes stochastic energy $E$ for each path. The two theory MLPs which are compared with data in Fig.~\ref{fig4} are shown superposed in black in panel (b); these are the paths which two distinct points on the manifold trace out as it evolves and folds over itself.
 }\label{fig3}
\end{figure*}

We now turn to the experimental investigation of the most likely path. To experimentally determine the most likely path we examine the Euclidean distance $d_i \equiv \frac{1}{M-1}\sum_{j\neq i}^M \sum_{k=0}^{N-1} \left((x_{i,k} - x_{j,k})^2 + (z_{i,k} - z_{j,k})^2\right)$ between each trajectory and all other trajectories in the ensemble.  Because the MLP should capture the highest density of other trajectories in a nearby vicinity, trajectories that minimize the Euclidean distance to others in the set should closely approximate the MLP. As such, we rank the trajectories in order of increasing $d_i$ and average the top few percent. This method was used to compare experiment and theory in the case of continuous quantum non-demolition measurement \cite{webe14}. From an ensemble of $\sim 10^5$ trajectories initialized in the excited state ($z=-0.97$), we post-select a sub-ensemble of trajectories that achieve a final given boundary condition, $\{z,x\}=\{ -0.3,-0.6\}$ at $t=1.94\ \mu$s, within a selection tolerance of $\pm 0.05$ as shown for $z$ in Fig.\ \ref{fig2}c. These trajectories are ranked according to the minimum Euclidean distance from all other trajectories and of these we highlight the closest $5\%$.  As shown in Fig.\ \ref{fig2}d, the average of these selected trajectories is in good agreement with the theoretical MLP.
\par An alternate method to determine MLP is to simply calculate the total path probability density (via Eq.\ \ref{joint_prob}) and average the top 5\% of trajectories with the highest path probability density. This naturally approximates the MLP.  In Fig.\ \ref{fig6} we compare these two methods and find that both methods to produce experimental MLPs that are in close agreement with theory. In contrast, a simple average of all of the pre- and post-selected trajectories is in clear disagreement with the theoretical MLP.

%Now, we turn to the experimental investigation of the most likely path. %The key ingredient of extracting the MLP out of an ensemble trajectories is the path 
%From an ensemble of $\sim 10^5$ trajectories initialized in the excited state ($z=-0.97$), we post-select a sub-ensemble of trajectories that achieve a final given boundary condition, $\{z,x\}=\{ -0.3,-0.6\}$ at $t=1.94\ \mu$s, within a selection tolerance of $\pm 0.05$ as shown for $z$ in Fig.\ \ref{fig2}c. These trajectories are ranked according to the minimum Euclidean distance from all other trajectories and of these we highlight the closest $5\%$.  As shown in Fig.\ \ref{fig2}d, the average of these selected trajectories is in good agreement with the theoretical MLP. 
\begin{figure}[t]
  \begin{center}
    \includegraphics[width=0.45\textwidth]{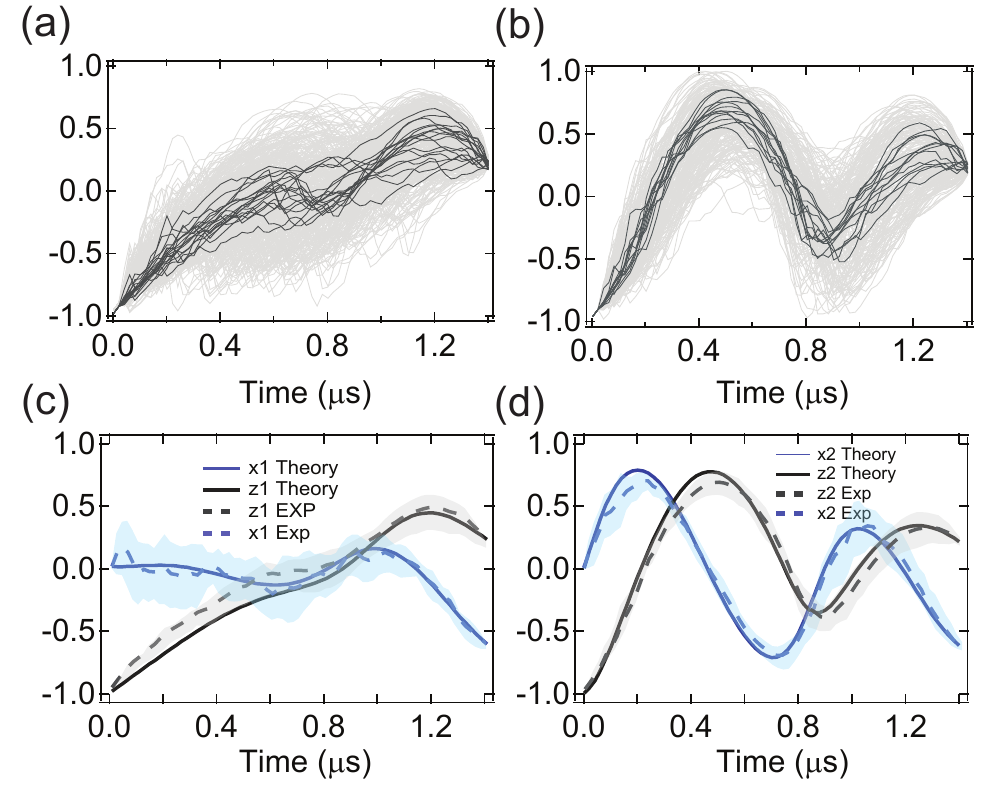}
  \end{center}
  \vspace{-.2in}
  \caption{\footnotesize \textbf{Experimental multiple most likely paths.} (a,b), By applying a clustering algorithm to experimental trajectories that attain the chosen initial and final boundary conditions we obtain two groups that minimize the mean Euclidean distance between members of each group.  The trajectories in each group with lowest Euclidean distance from other members in the group are highlighted in black ($z$ trajectories are shown).  (c,d), Comparison to theory; the solid curves are theory predictions and the dashed lines are obtained from averaging the highlighted experimental trajectories. The shaded area along experimental curves shows the standard deviation of the averaged trajectories.}\label{fig4}
\end{figure}

%\section{Multiple Most likely path}
\section{Multiple most likely paths}

The MLP is the solution for the equations-of-motion, Eqs. (\ref{eq-mlp2witheta}). Now, it is natural to ask if these  equations-of-motion have more than one solution as suggested from the presence of caustics in ray optics. Multiple solutions exist at states where a Lagragian manifold in MLP phase space overlaps itself, either through a fold or divergence of the manifold \cite{Littlejohn1992,Dykman1994_LM}, or at a ``winding number'' overlap caused by the manifold wrapping over itself as constituent paths orbit the Bloch sphere at different speeds \cite{phil2016}. In either case, we expect distinct clusters of stochastic trajectories corresponding to the different MLPs. As shown in Fig. \ref{fig3}, we may find such solutions theoretically, in Eqs. (\ref{eq-mlp2witheta}), by choosing an initial state coordinate ${\bf q_i}$, and sweeping through different initial momenta ${\bf p_i}$; this defines the Lagrangian manifold we use throughout the subsequent analysis. Evolution under Eqs. (\ref{eq-mlp2witheta}) deforms this initially flat, vertical plane in MLP phase space because different initial momenta result in different final states. When the manifold %fails the vertical line test at a state and elapsed time (
is no longer a one-to-one function between $\mathbf{q}$ and $\mathbf{p}$%)
, multiple paths corresponding to extrema in the stochastic action connect the initial and final states.

We note that in the absence of drive,  the evolution of the qubit is constrained to a deterministic ellipse for a given evolution time \cite{ibarcq2015,nagh16,dian17}. In this case, the Lagrange Manifold is constrained to these  deterministic ellipses.  As a result the  manifold is always a one-to-one function between $\mathbf{q}$ and $\mathbf{p}$, and multiple-MLPs are not possible.  As seen in Fig.~\ref{fig3}b,c the drive breaks this one-to-one character leading to an increasing number of MLPs for longer evolution times.

In order to confirm the presence of these multiple-MLPs in experimental data, we pre- and post-select the trajectories by the given boundary conditions. While the experimental trajectories do not exactly follow one or the other MLP, the MLP solutions should approximate the paths taken by many of the individual trajectories.  In this case, the pre- and post-selected trajectories should belong to two different groups, such that the mean Euclidean distance between the members in each group is minimized. Therefore, we need a clustering algorithm \cite{reic06,newm06,fort10} that efficiently separates these trajectories into two groups.
%\subsection{Clustering algorithm}
%To compare experimental data to theory for boundary conditions that yield two solutions to Eqs.\ (6) we post-select on trajectories that meet these same boundary conditions with a tolerance of $\pm 0.05$. While these trajectories are stochastic, they are likely to follow one of the two deterministic most likely paths.
  In order to perform clustering, for each $N$-step trajectory we  define the weighted mean Euclidean distance to other elements of the set $d^W_i \equiv \frac{1}{M-1}\sum_{j\neq i}^M \sum_{k=0}^{N-1} P_j \left((x_{i,k} - x_{j,k})^2 + (z_{i,k} - z_{j,k})^2\right)$ and define the average weighted Euclidean distance of the set as $\bar{d}^W = \frac{1}{M}\sum_{i=1}^M d_i^W$, where $M$ is the number of trajectories in the set and $P_j$ is a weighting factor that is proportional to the logarithm of the probability for trajectory $j$ as determined from Eq.\ \ref{joint_prob}.  The algorithm aims to separate the ensemble into two clusters that minimize the sum of the weighted Euclidean distance between members of each set, $\bar{d}^W_\mathrm{set1}+ \bar{d}^W_\mathrm{set2}$.  The clustering algorithm starts by randomly splitting the ensemble of trajectories into two sets. In each iteration of the algorithm the weighted distance, $d^W_i$, given a randomly chosen trajectory $i$, is calculated for each set and the trajectory $i$ is added to the set that minimizes $\bar{d}^W_\mathrm{set1}+ \bar{d}^W_\mathrm{set2}$. The algorithm proceeds through the ensemble of trajectories by transferring trajectories between clusters to minimize the average weighted distance.  We find that the algorithm typically finds  the  optimal configuration  in  $\simeq M$ steps and finds the same clustering configuration independent of the initial random configuration (see Figure \ref{fig4}a,b). The weighting factor $P_j$ helps the algorithm converge to the final configuration faster, but does not substantially affect the final cluster distributions.
\par By averaging the individual paths of each group that have the minimum Euclidean distance from other members in each set,  we obtain the experimental most likely paths (\ref{fig4}c,d). These experimental MLPs are in reasonable agreement with the two MLPs terminating at the desired boundary conditions. % I am not sure we need the comment after the semi-colon at all actually (PL).

While the occurrence of multiple most likely paths is associated with multiple extrema of the stochastic action, these solutions may have different values of stochastic action---meaning that the paths have different total path probability densities.  We can check if these path probability densites are accurately represented by the relative occurrence of paths in each cluster.
 
%we analyze four different boundary conditions leading to multiple MLPs and find that the predicted relative probability densities are in agreement with the trajectory distributions obtained through clustering.
\begin{figure}[]
  \begin{center}
    \includegraphics[width=0.45\textwidth]{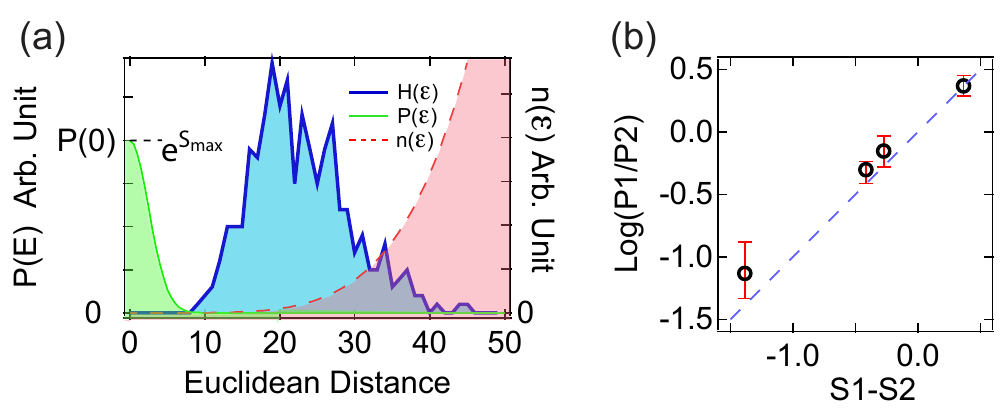}
  \end{center}
  \vspace{-.2in}
  \caption{\small  \textbf{Relative probability of multiple most likely paths.} (a), The peak of the probability density distribution of the Euclidean distance from the MLP (green) is related to the extremum of the stochastic action ($\mathcal{S}_\mathrm{max}$).  The observed distribution (blue) has a maximum that shifted from $0$ due to the increasing multiplicity of trajectories with a larger Euclidean distance from the MLP. (b), To compare the experimental distributions to the theoretical stochastic entropy maxima, we use the multiplicity ($n(\mathcal{E})$) and observed distribution $H(\mathcal{E})$ to determine $P(0)$.  We compare the relative probability  $(P(0)_\mathrm{cluster1}/P(0)_\mathrm{cluster2})$ to the predicted relative probability $S_1-S_2$.    Error bars indicate the uncertainty in the fit to the relative distribution. We apply this analysis to 4 different MMLP pairs with initial state $\{z(0),x(0)\}=\{-0.97,0\}$ and final states 
 $\{z(T),x(T),T(\mu $s$) \} = \{0.65,-0.08,1.2  \}$, $\{0.19,-0.93,1.4 \}$, $\{0.36,0.47,1 \}$, $\{0.21,-0.62,1.4\}$, (left to right on graph). }\label{subfig1}
\end{figure}
%\subsection{Relative probability of multiple most likely paths}
%While multiple most-likely paths occur when there are two maxima to the stochastic action $\mathcal{S}$, these maxima in general do not attain the same value. This means that one path is more probable than the other.
 Figure \ref{subfig1}a displays a histogram of the Euclidean distance, $\mathcal{E}$, of trajectories from their MLP in each set. Given the fact that the MLP captures most trajectories in its vicinity, one would think these histograms would acquire a maximum at $\mathcal{E}=0$ and decrease as we go far from the MLP. However, the multiplicity increases for trajectories with larger $\mathcal{E}$ and the peak of the histogram occurs for non-zero $\mathcal{E}$.  The multiplicity, $n(\mathcal{E})$ which  accounts for the number of different trajectories that have same Euclidean distance $\mathcal{E}$ from a MLP is given by the number of ways a Euclidean distance can be obtained from $N$ timesteps, $\mathcal{E}_1^2+\mathcal{E}_2^2+...+\mathcal{E}_N^2=\mathcal{E}^2$, where $\mathcal{E}_i$ is the deviation from the MLP at the $i^\mathrm{th}$ time step.

With this understanding of the multiplicity, the distribution of trajectories attaining an average Euclidean distance from the MLP, $H(\mathcal{E})$, is given by the product of the path probability distribution and the multiplicity $H(\mathcal{E}) = P(\mathcal{E}) \times n(\mathcal{E})$, 
\begin{align}
H(\mathcal{E})= P(0) e^{- \frac{\mathcal{E}^2}{2 \sigma^2}} \times \frac{2 \pi^{N/2}}{\Gamma(\frac{N}{2})} \mathcal{E}^{N-1},
\end{align} 
where we assume that $P(\mathcal{E})$ is  a half-normal distribution  centered at zero and characterized by the variance $\sigma^2$. The multiplicity $n(\mathcal{E})$ is given by the area of an $N$-dimensional hypersphere, and $\Gamma$ is the Gamma function.
The variance $\sigma^2$ can be determined directly, given the fact that $H(E)$ acquires its maximum at $\mathcal{E}=\sigma\sqrt{N-1}$. Since we are interested in  the relative probability of two MLPs, $P_1(0)/P_2(0)$ we divide two experimental histograms,
\begin{eqnarray}
\frac{H_1(\mathcal{E})}{H_2(\mathcal{E})} = \frac{P_1(0)}{P_2(0)} \exp\bigg[- \frac{\mathcal{E}^2}{2}\bigg( \frac{1}{\sigma_1^2} - \frac{1}{\sigma_2^2} \bigg)\bigg], \label{eq:rel}
\end{eqnarray}
and fit this function to the data to extract the relative probability. We show the result in Fig.\ \ref{subfig1} which has reasonable agreement with the theoretical path probability values calculated by stochastic action.

%\par {\color{cyan} We make a few qualitative comments about the two MLPs in Figs. 4 and 5; we can see that they come into being through the competition of different phenomena in the experiment, namely the Rabi drive, decay due to the fluorescence, and (imperfectly-efficient) measurement. One path orbits the sphere with the Rabi drive, losing purity at an approximately steady rate; we infer that the drive is the primary generator of these dynamics, with measurement inefficiency steadily eroding the purity. But a second MLP exists with reasonably high probability, which decays into mixed states very quickly (resisting the drive as it does), and then circles around with the drive for a short time inside the Bloch sphere, to arrive at the same final state, with one winding count less (some discussion of winding numbers can be found in the appendices, and in Ref. \cite{phil2016}). This second path, in other words, starts out behaving like we expect paths to behave without the drive at all \cite{nagh16}, generated only by a continuous measurement of the fluorescence signal, and then gets ``caught'' by the drive after the state becomes quite mixed.}

Looking at the two MLPs shown in Figs. 4 and 5, we immediately see that they have different winding numbers about the Bloch sphere (see the appendices, and/or Ref. \cite{phil2016} for a discussion of winding number MMLPs). A given noise realization or measurement record can move the state approximately with the Rabi drive, or move the state substantially ahead or behind the drive; thus it is possible to obtain MLPs which end at the same state, having gone around the Bloch sphere a different number of times. What we have with the MMLP example shown is (1) a path which moves approximately with the drive, and (2) a path whose initial momentum corresponds to an optimal readout which lags behind the drive, such that it reaches the given final state with one less winding count. %Since initially pure states remain pure for perfect measurement efficiency $\eta = 1$, we may suppose that the measurement inefficiency is the primary cause of the states becoming mixed for both paths. 
%They respond to this effect quite differently however; we may infer that the path which becomes quickly mixed is responding immediately to measurement back action and very little to the drive at first, before getting caught by the drive once the state is already very mixed. The other path is immediately caught by the drive, and follows it for its whole evolution, losing purity at a much more steady rate. 
In summary, quantum dynamics of the MLP is given by the competition between the stochastic evolution associated with measurement and the unitary evolution arising from the drive.  Different MLPs occur when one or the other types of evolution dominate at different times.
This example illustrates that the existence of several MLPs arises from the presence of several different effects in the system; they can each take precedence over the other, in a given case, and generate qualitatively different behaviors that land at the same final state.

\section{Conclusion}
In the absence of measurement, a closed quantum system has a dynamics that is described by the Schr\"odinger equation.  Given the initial condition, a {\it unique, deterministic} solution emerges that specifies the quantum state at all future times. In contrast, resonance fluorescence is an intrinsically open quantum system phenomenon, leading to a dynamics that differs dramatically from that of a closed system. This difference is highlighted in the phenomenon of {\it quantum caustics};  even for relatively short periods of evolution, catastrophes may form in manifolds of the most-likely path phase space, generating caustic regions where several most-likely paths link a given initial and final state over a given time evolution.
%causing caustic regions to come into existence, which signal the onset of multiple most likely paths between the same fixed initial and final state boundary conditions. 
By understanding multiple most likely paths through this catastrophe formation in the manifold, %the folding of a Lagrange manifold in the most likely path phase space, 
we see an analogy between the multiple-path propagation of our quantum trajectories, and caustic phenomena in optical propagation through random media.
These phenomena may have important consequences for quantum control, and opens new possibilities in the investigation of dynamical instabilities and chaos in continuously measured quantum systems.
% PL: Do we want to cite any quantum control / chaos sources that back this up, or just leave it hanging for the moment?

\section*{Acknowledgment}
We acknowledge helpful discussions and early contributions from A. Chantasri. We acknowledge primary research support from the John Templeton Foundation grant ID 58558, NSF grants DMR-1506081 and PHY-1607156, and secondary support for personnel from the ONR No. 12114811 and the ARO No. W911NF-15-1-0496 . This research used facilities at the Institute of Materials Science and Engineering at Washington University. K.W.M acknowledges support from the Sloan Foundation.

%\bibliographystyle{unsrt}
%\bibliography{bibfile}

%\section{Supporting Information (SI)}
\section{Appendix: Reduction of the Stochastic Hamiltonian} \label{ap_pure}
We may reduce our stochastic Hamiltonian Equ.\ \ref{h4} to an even simpler form. Consider a canonical transformation from Cartesian to polar coordinates,
\be \label{convert4}  \left( \begin{array}{c} x \\ z \\ p_x \\ p_z \end{array} \right ) =  \left( \begin{array}{c} R\sin\theta \\ R\cos\theta \\ p_R \sin\theta + p_\theta \cos\theta / R \\p_R \cos\theta - p_\theta \sin \theta / R \end{array} \right ), \ee
where the Poisson brackets $\lbrace x, p_x \rbrace  = 1 = \lbrace z, p_z \rbrace$ are preserved to obtain $\lbrace R, p_R \rbrace = 1 = \lbrace \theta, p_\theta \rbrace$. We %let $u \rightarrow 1+z$, and 
apply the transformation \eqref{convert4} to \eqref{h4} to obtain $H \rightarrow \mathfrak{h}$. We may once again obtain equations of motion from Hamilton's equations in the new coordinates, \emph{i.e.} $\dot{R} = \partial_{p_R} \mathfrak{h}$, $\dot{\theta} = \partial_{p_\theta} \mathfrak{h}$. We find that
\be
\dot{R}|_{R=1,\eta=1} = 0,
\ee
and that $\dot{\theta}$ does not depend on $p_R$, meaning that the phase-space can be reduced further to two dimensions, by restricting ourselves to pure states ($R=1$), which stay pure with perfect measurement efficiency ($\eta=1$), in the $xz$-plane; states are then parameterized entirely by the polar angle $\theta$. No experimental detection scheme can presently achieve $\eta = 1$, but this idealized case is still useful to gain insight into the behaviors we see in more realistic forms of the system, while keeping the mathematics as simple as possible.
%\subsection{Analysis of 2D Phase-Space}
\par The stochastic Hamiltonian for the two-dimensional phase space is $h = \mathfrak{h}|_{R=1,\eta=1}$, or
\be\begin{split}
h =& p\left( -\Omega + r\sqrt{\gamma}(1-\cos\theta ) - \gamma \sin\theta \right) \\&- \sqrt{\gamma} r \sin\theta - \frac{r^2+\gamma(1-\cos\theta)}{2}
\end{split}\ee
where $p_\theta \rightarrow p$. The optimal readout for the MLPs is $r^\star = -\sqrt{\gamma}(p (\cos\theta-1)+\sin\theta)$, and is obtained by solving $\partial_r h|_{r=r^\star} = 0$ for $r^\star$.

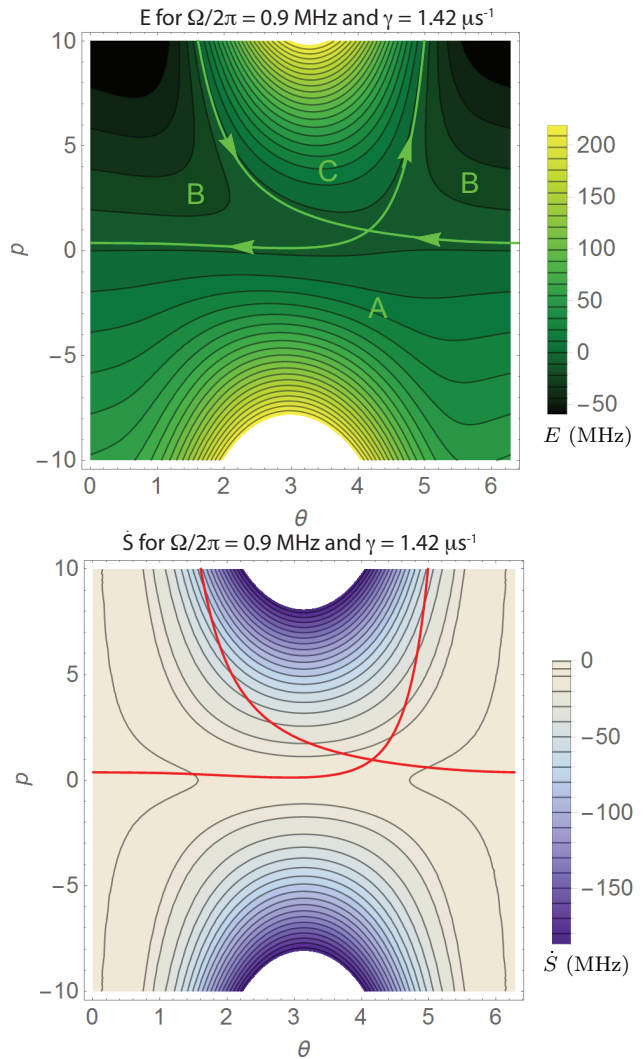
\begin{figure}
%\begin{tabular}{c}
%$E$ for $\Omega/2\pi = 0.9$ MHz and $\gamma = 1.42 \mu s^{-1}$ \\
%\includegraphics[width=.98\columnwidth]{2D_flor_ps_ex.pdf}
\includegraphics[width=.98\columnwidth]{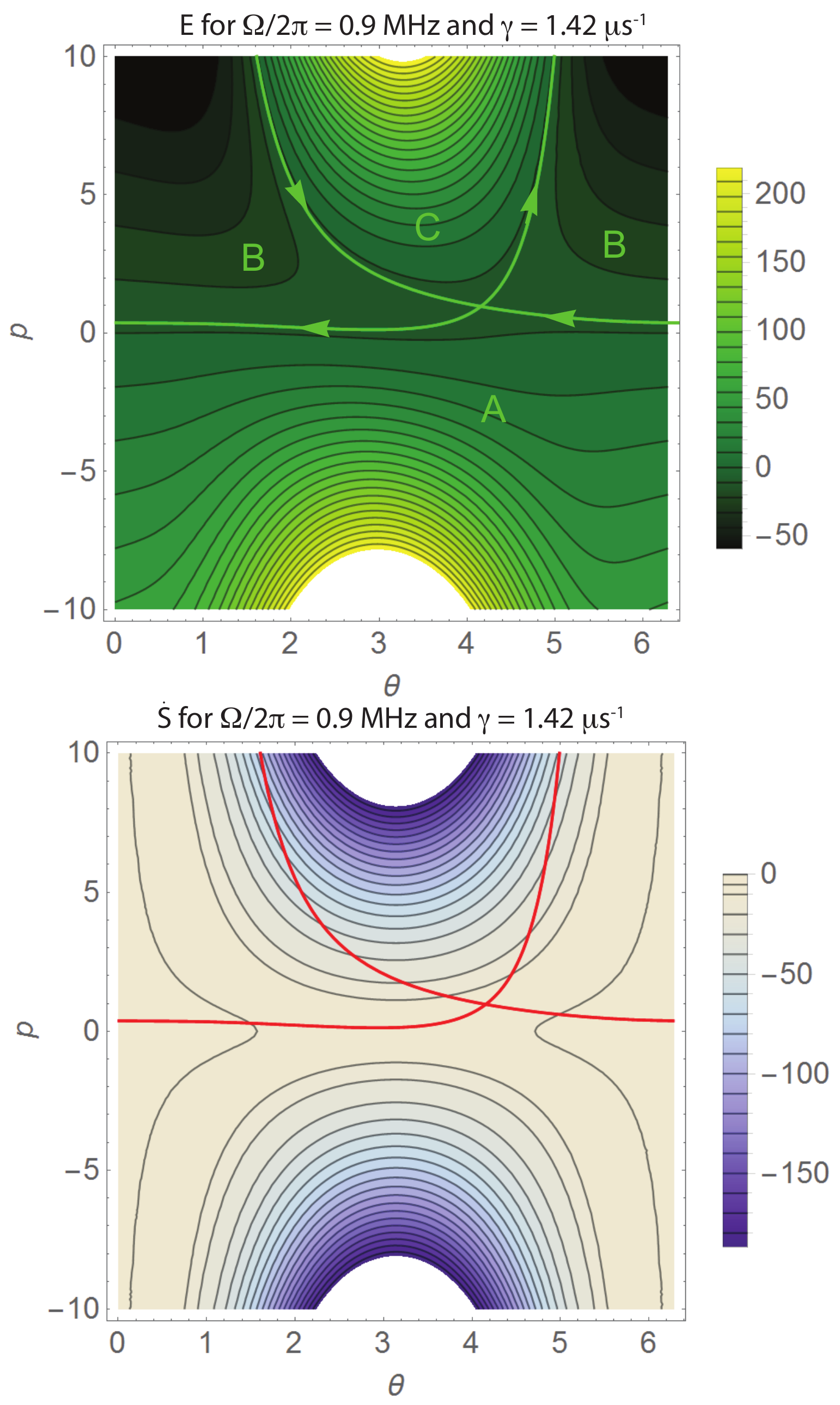}
\begin{picture}(1,1)
\put(-40,35){\footnotesize $\dot{S}$ (MHz)}
\put(-40,235){\footnotesize $E$ (MHz)}
\end{picture}
%\\ 
%$\dot{S}$ for $\Omega/2\pi = 0.9$ MHz and $\gamma = 1.42 \mu s^{-1}$ \\
%\includegraphics[width=.98\columnwidth]{2D_flor_sd_ex.pdf}
%\end{tabular}
\caption{We show lines of constant stochastic energy (top), and lines of constant $\dot{S}$ (bottom). $E$ and $\dot{S}$ are in the same units of MHz as $\Omega$ and $\gamma$. Green (top) and red (bottom) lines are those for the energy $E_c  =-2.13$ MHz containing the lone fixed point at $(\bar{\theta} = 4.16, \bar{p} =0.967 )$. Paths in region A all travel right to left (with the Rabi drive), and remain relatively probable over long periods of time provided they have a modest momentum/stochastic energy. Paths in region B contain both branches $p_+$ and $p_-$ \eqref{p_pm}, which move in opposite directions; these paths may also have reasonably high relative probabilities to occur for initial conditions just to the left of the fixed point, over modest time evolutions (approximately less than the time required to almost a period/approach the next fixed point), but asymptote into regions of very negative $\dot{S}$ over long time intervals. Paths in region C only go left to right (against the Rabi drive), and inhabit regions of more negative $\dot{S}$, meaning that they are relatively improbable over any appreciable time interval. }\label{fig-ps1sd1}
\end{figure}

We additionally write the stochastic Hamiltonian with the optimal readout substituted in, which reads $h^\star = a(\theta) p^2 + b(\theta) p + c(\theta)$, where we have
\begin{subequations}\label{abc-eqmot}
\begin{eqnarray}
{a(\theta)}  &=&  \gamma \left( -\cos\theta + \frac{1 + \cos^2\theta}{2} \right), \\
{b(\theta)}  &=& -\Omega + \frac{\gamma}{2} \left(-3 \sin \theta + \sin(2\theta) \right), \text{ and} \\
{c(\theta)} &=& -\frac{\gamma}{2} \left(\cos^2\theta - \cos\theta \right).
\end{eqnarray}  
\end{subequations}
The ``probability cost-function,'' discussed in related work \cite{phil2016} is
\be
\dot{S} = \gamma \sin ^2\left(\frac{\theta }{2}\right) \left(p^2 (\cos\theta - 1) + \cos\theta\right),
\ee
which is the integrand of the stochastic action $\mathcal{S}$, and describes approximately how the probability density for a MLP is affected by time spent in different regions of phase space.  (The full joint probability may be written $\mathcal{P}\sim e^\mathcal{S}$ in the small noise approximation.) Note that $h^\star = p \dot{\theta}+ \dot{S}$, and therefore $\dot{S} = - ap^2 + c$.
We plot the phase-portrait and contour plot of $\dot{S}$ in Fig.~\ref{fig-ps1sd1}. The phase-portrait contains lines of constant stochastic energy $E = h^\star$, which may be solved to obtain functions $p_\pm(\theta,E)$. These have the form
\be \label{p_pm}
p_\pm(\theta,E) = -\frac{b}{2a} \pm \sqrt{\frac{E-c}{a}+ \frac{b^2}{4 a^2}},
\ee
where $a$, $b$, and $c$ are shown in \eqref{abc-eqmot}. Plots of the lines of constant stochastic energy (phase-portrait) and lines of constant $\dot{S}$ are included in Fig.~\ref{fig-ps1sd1}. 

\definecolor{wgreen}{RGB}{51,255,153}
\begin{figure}
\begin{tabular}{cc}
\includegraphics[height=.8\columnwidth]{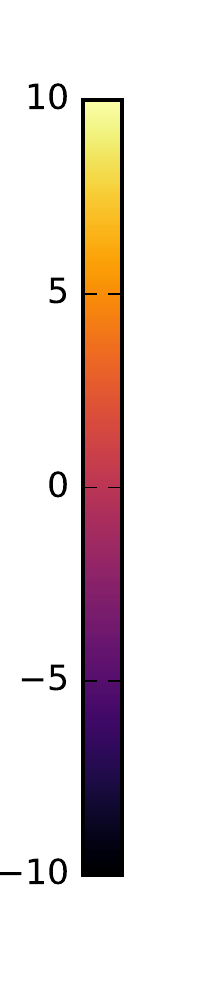} & \includegraphics[height=.8\columnwidth]{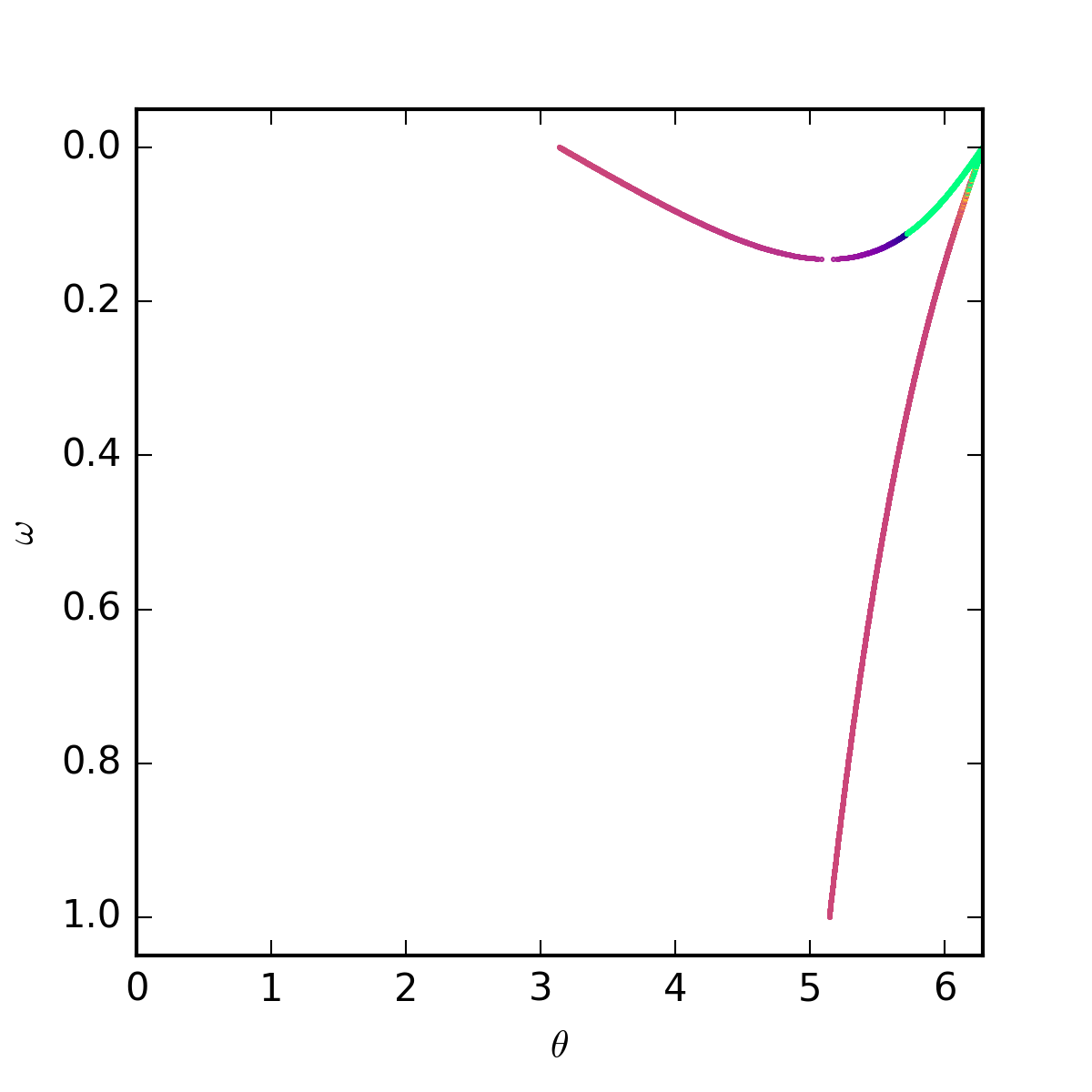} 
\end{tabular}
\begin{picture}(1,1)
\put(-115,16){\color{wgreen} $p < -10$}
\put(-115,193){\color{wgreen} $p > 10$}
\end{picture}
\caption{We plot the locations of fixed points $\bar{\theta}$ against the dimensionless parameter $\omega = \Omega/\gamma$. Color denotes momentum $\bar{p}$. Experimentally we operate at $\omega = 0.63 \cdot 2\pi$, where there is only one fixed point. This fixed point is a location in the phase space where the Rabi drive  and fluorescence cancel. Another pair forms for $\omega < 0.145 = \omega_c$, splitting around $\theta = 5.18$. As the drive is eliminated entirely, these show system is able to stall at either the ground $(\theta = 0$ or $2\pi)$ or excited $(\theta=\pi)$ states. All fixed points exist on the half of the Bloch sphere where the fluorescence and drive compete (the drive pushes the state from ground to excited, whereas the fluorescence pushes the state from excited to ground); none exist on the half where the fluorescence and drive work together (both push towards the ground state), indicating that no optimal readout can fight both the drive and spontaneous emission at the same time to hold the system still at some particular state.} \label{fig-fixpoint}
\end{figure}
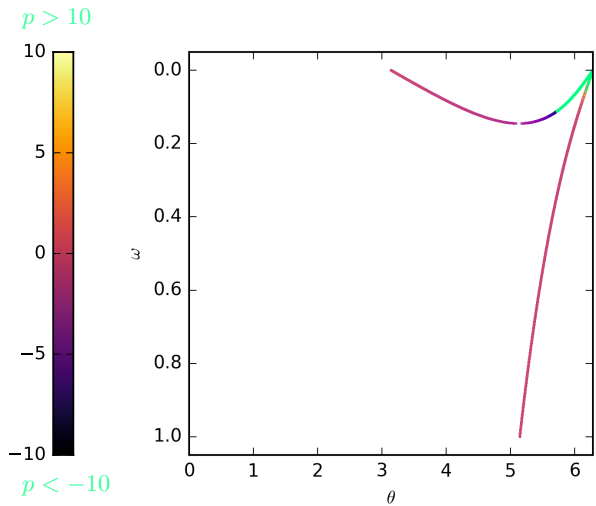

\par Hamilton's equations are a subset of dynamical systems of the form $\dot{\mathbf{x}} = \mathbf{F}[\mathbf{x}]$, where here $\mathbf{x}  = (\theta,p)$. Fixed points of a dynamical system are points $\bar{\mathbf{x}}$ which satisfy $\mathbf{F}[\bar{\mathbf{x}}]= 0$, \emph{i.e.} denote points where the system sits still. We search for the fixed points $(\bar{\theta},\bar{p})$ of this system, noting that 
\begin{subequations}\label{eqmot}
\begin{eqnarray}
{\dot{\theta}}  &=&  2 a p + b, \\
{\dot{p}}  &=& -a'p^2 - b' p - c', 
\end{eqnarray}  
\end{subequations}
where $a$, $b$, and $c$ are functions defined in \eqref{abc-eqmot}, and $a'$, $b'$, and $c'$ are their derivatives with respect to $\theta$. Then $\dot{\theta} = 0$ leads to $\bar{p} = -b/2a$, which may be put into the expression for $\dot{p} = 0$ to obtain
\be\label{fixpoint_condition}
a' \left(\frac{b}{2a}\right)^2 - b' \left(\frac{b}{2a}\right) + c' =0.
\ee
Any fixed-point coordinate $\bar{\theta}$ must satisfy \eqref{fixpoint_condition}. The LHS of \eqref{fixpoint_condition} depends only on $\theta$ and the dimensionless ratio $\omega \equiv \Omega/\gamma$. We find that \eqref{fixpoint_condition} admits one solution for $\omega > 0.145$, and three solutions for $\omega < 0.145$, shown diagrammatically in Fig.~\ref{fig-fixpoint}. We define a critical $\omega_c = 0.145$ where this bifurcation occurs. We may interpret this in terms of the optimal readout; we require that $\dot{\theta}= -\Omega - \gamma \sin\theta + r \sqrt{\gamma}(1-\cos\theta)$ be zero, which is equivalent to a requirement that the terms for driving, fluorescence, and measurement back-action (written in that order), cancel out. The constraint on $p$ in the MLP phase space enters indirectly through the optimal readout. We may understand a fixed point as a combined value of $\theta$ and readout where the system will sit still.
\par The parameters used in the accompanying experiment are $\Omega/2\pi = 0.9$ MHz and $\gamma = 1.42\ (\mu \mathrm{s})^{-1}$, well above the value of $\omega$ where there are three fixed points. At these parameters there is only one fixed point, which sits on the separatrix of energy $E_c = -2.126$ MHz shown in Fig.~\ref{fig-ps1sd1}. Its location is $(\bar{\theta},\bar{p}) = (4.155,0.967)$.

\section{Appendix: Winding Number MMLP Similar to Experiment}
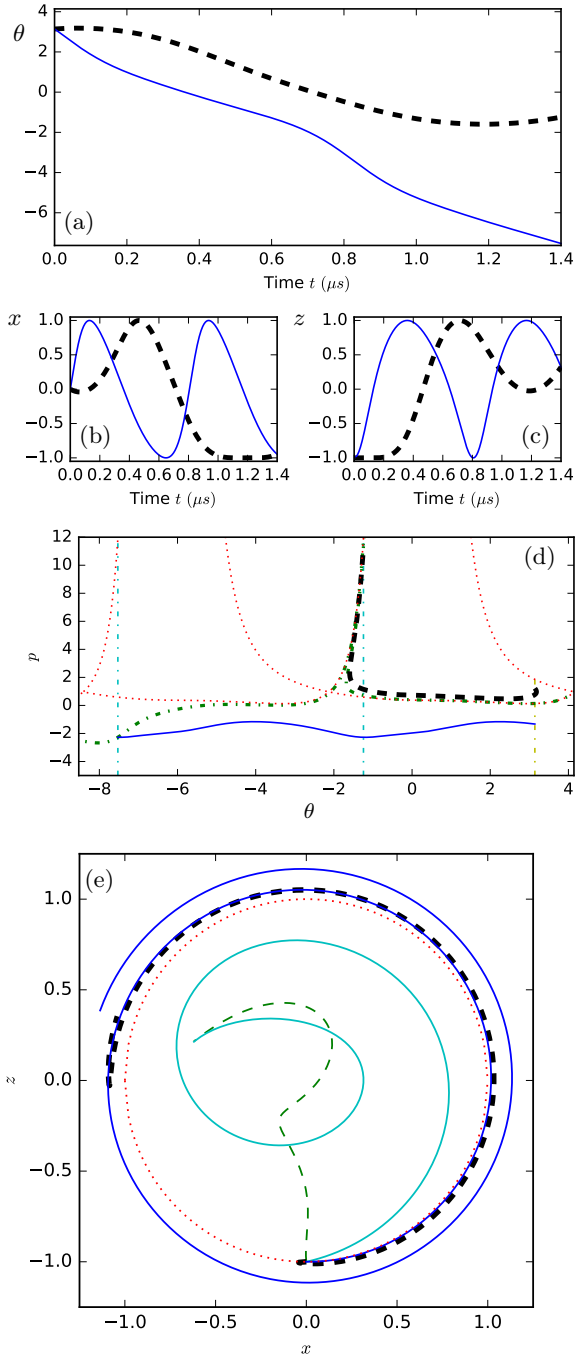
\begin{figure}
\begin{tabular}{c}
\includegraphics[width=.9\columnwidth]{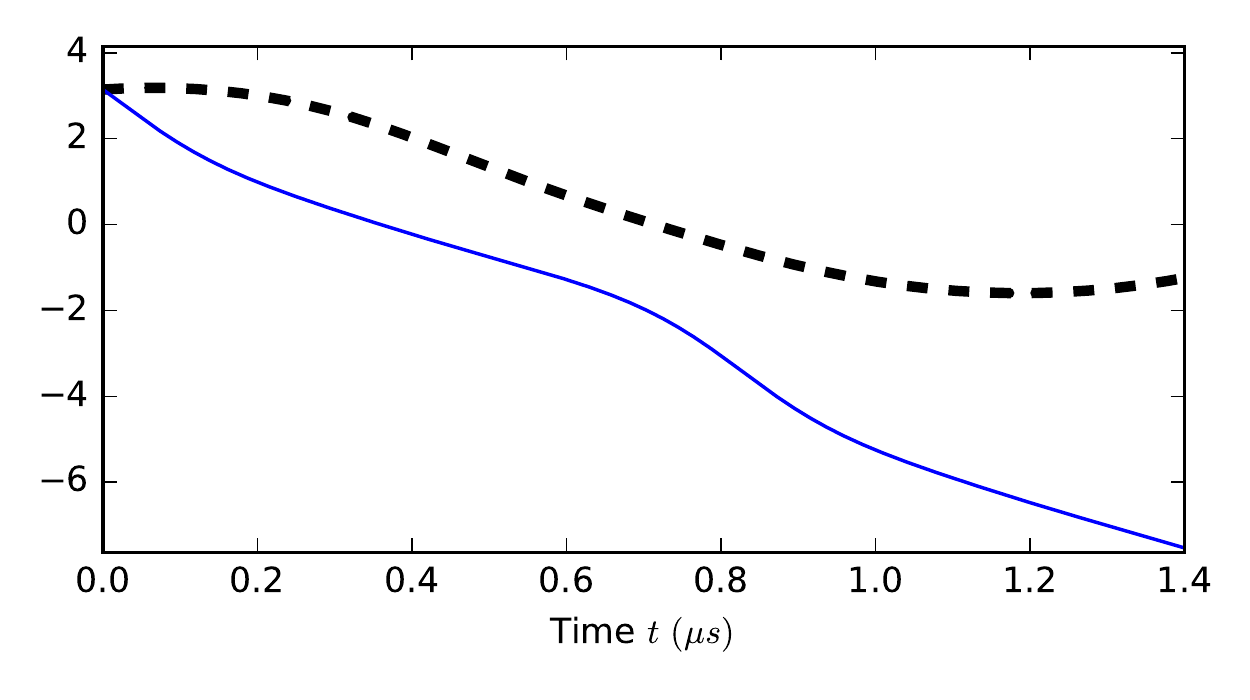} \vspace{-10pt} \\
\includegraphics[width=.9\columnwidth]{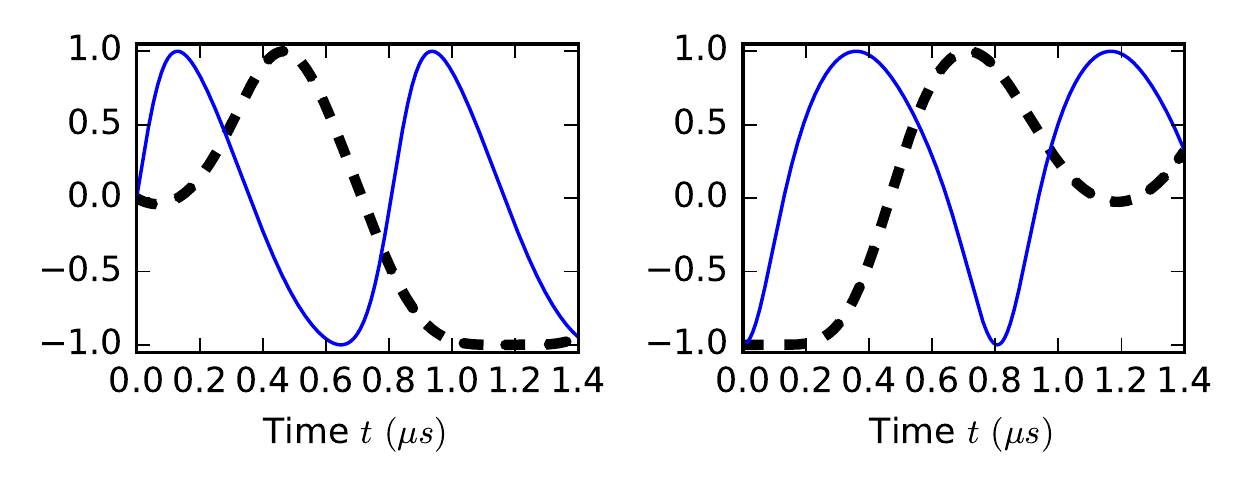} \vspace{-10pt} \\
\includegraphics[width=.9\columnwidth]{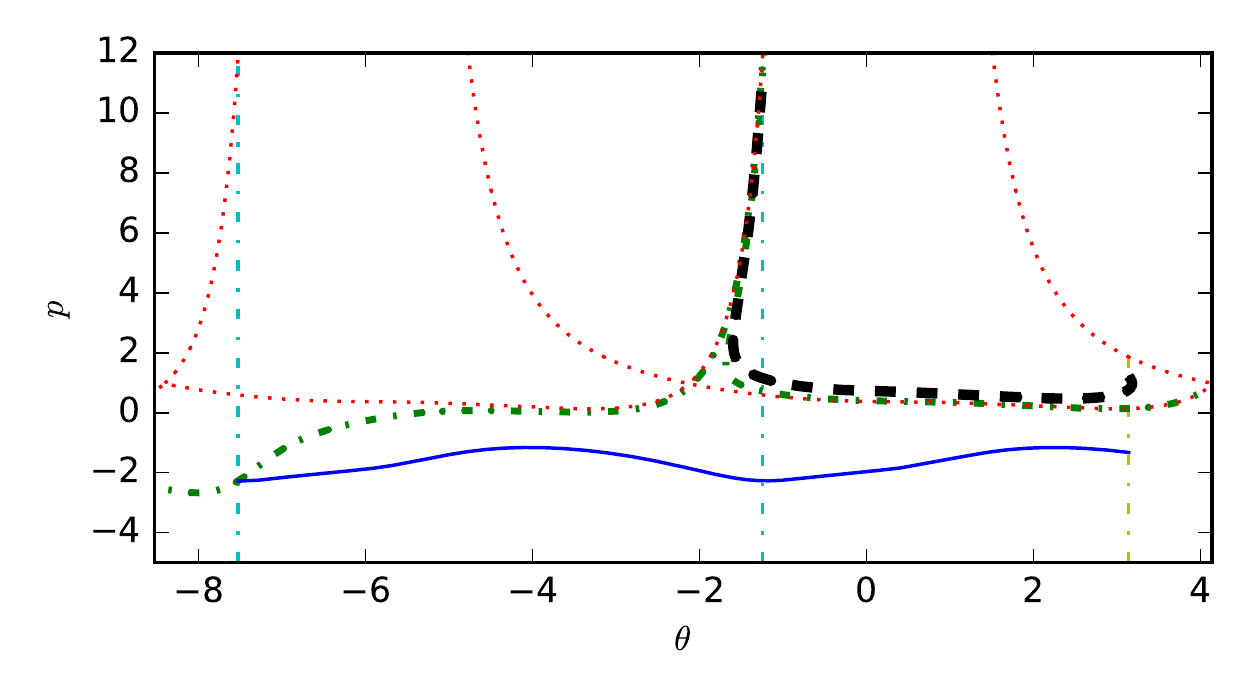} \vspace{-20pt} \\ 
\includegraphics[width=.9\columnwidth]{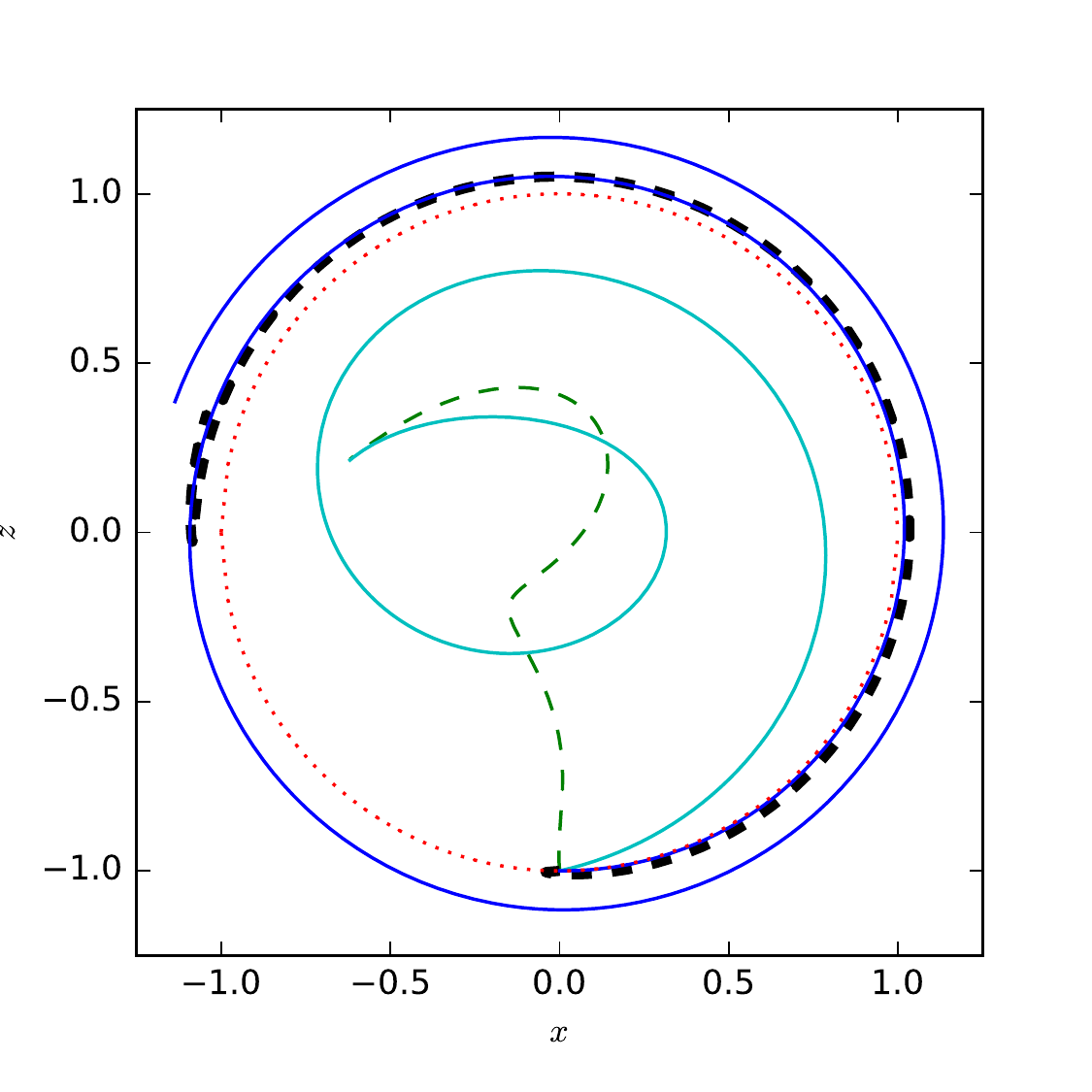} \vspace{-16pt}
\end{tabular} \begin{picture}(1,1)
\put(-227,239){$\theta$}
\put(-229,131.5){$x$}
\put(-121,131.5){$z$}
\put(-208,168){(a)}
\put(-202,87){(b)}
\put(-35,87){(c)}
\put(-34,41){(d)}
\put(-200,-81){(e)}
\put(-117,-55){\footnotesize $\theta$}
\end{picture}
\caption{We show MMLPs with different winding numbers, from $\theta_i = \pi$ to either $\theta_f^{(B)} = -1.24$ (dashed black) or $\theta_f^{(A)} = -1.24- 2\pi$ (solid blue). These are shown as $\theta(t)$ (a), $x(t)$ and $z(t)$ (b,c), and in phase space (d). In the phase space plot we show $\theta_i$ in dash-dotted yellow (the Lagrangian manifold at $t=0$), the two $\theta_f$ in dash-dotted cyan, the separatrix in dotted red, and the Lagrangian manifold at $T = 1.4\ \mu s$ in dash-dotted green. Finally, in (e) we show the pure state paths (still blue and dashed black) in the $xz$-plane of the Bloch sphere. The dotted red line marks the edge of the sphere; the pure-state curves are given an artificial radius outside the sphere for added visibility. The cyan and dashed green curves in (e) are the theory calculations for the example with $\eta = 0.45$ from Figs. 4 and 5 of the main text, shown for comparison.} \label{fig-wn}
\end{figure}

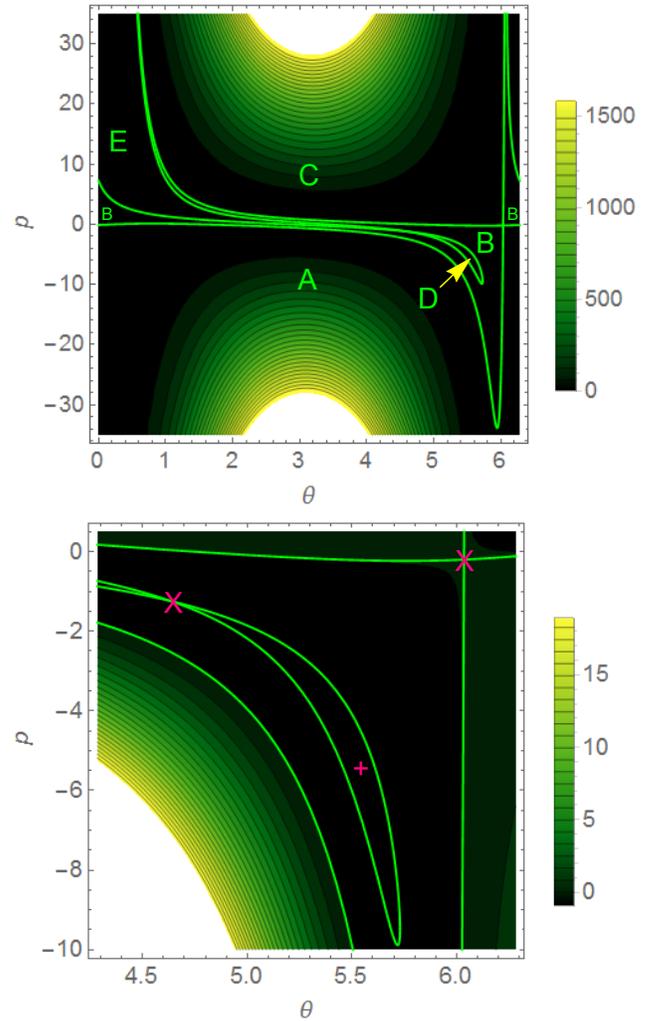
\begin{figure}
\begin{tabular}{l}
\includegraphics[width=.98\columnwidth]{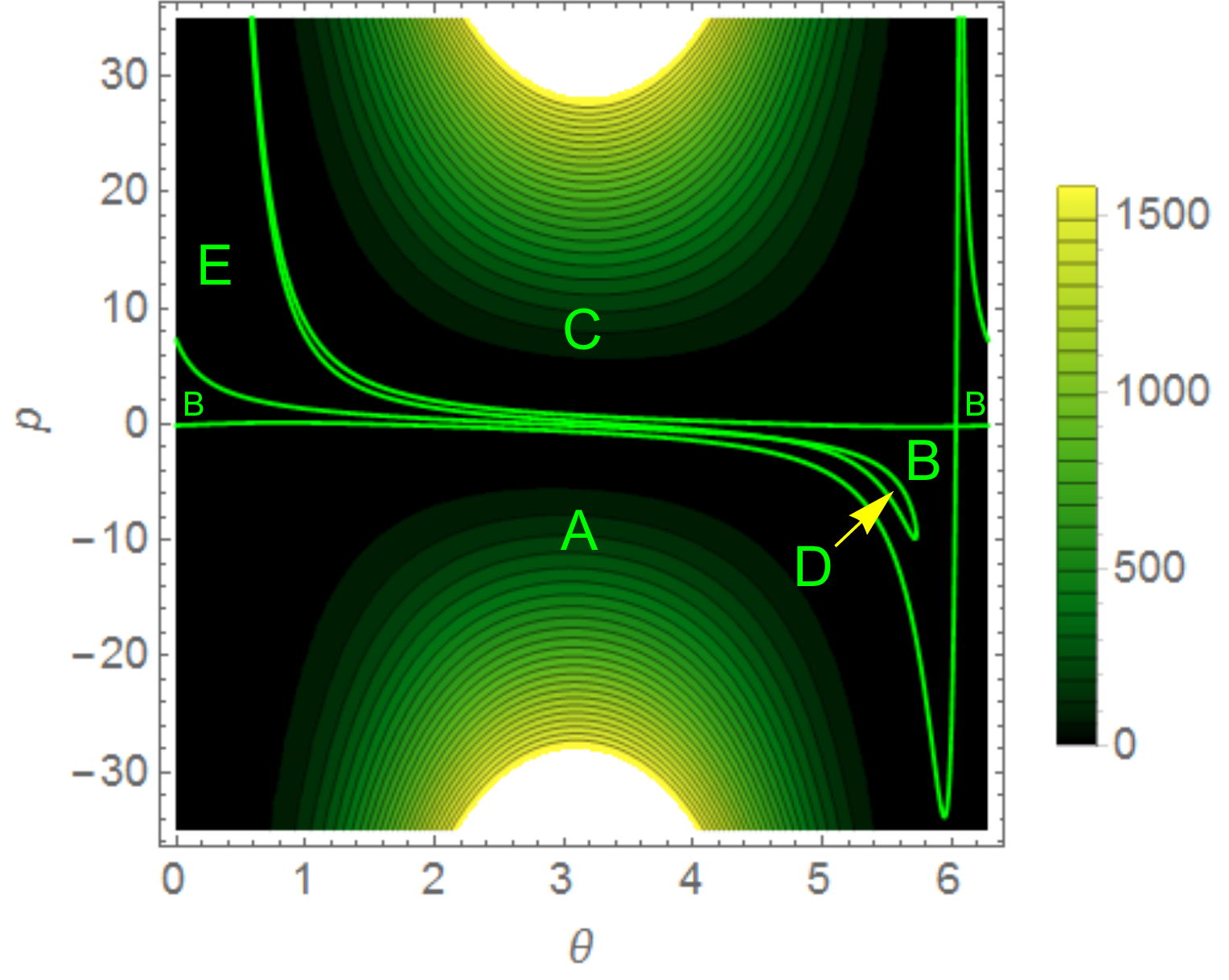} \\
\includegraphics[width=.94\columnwidth]{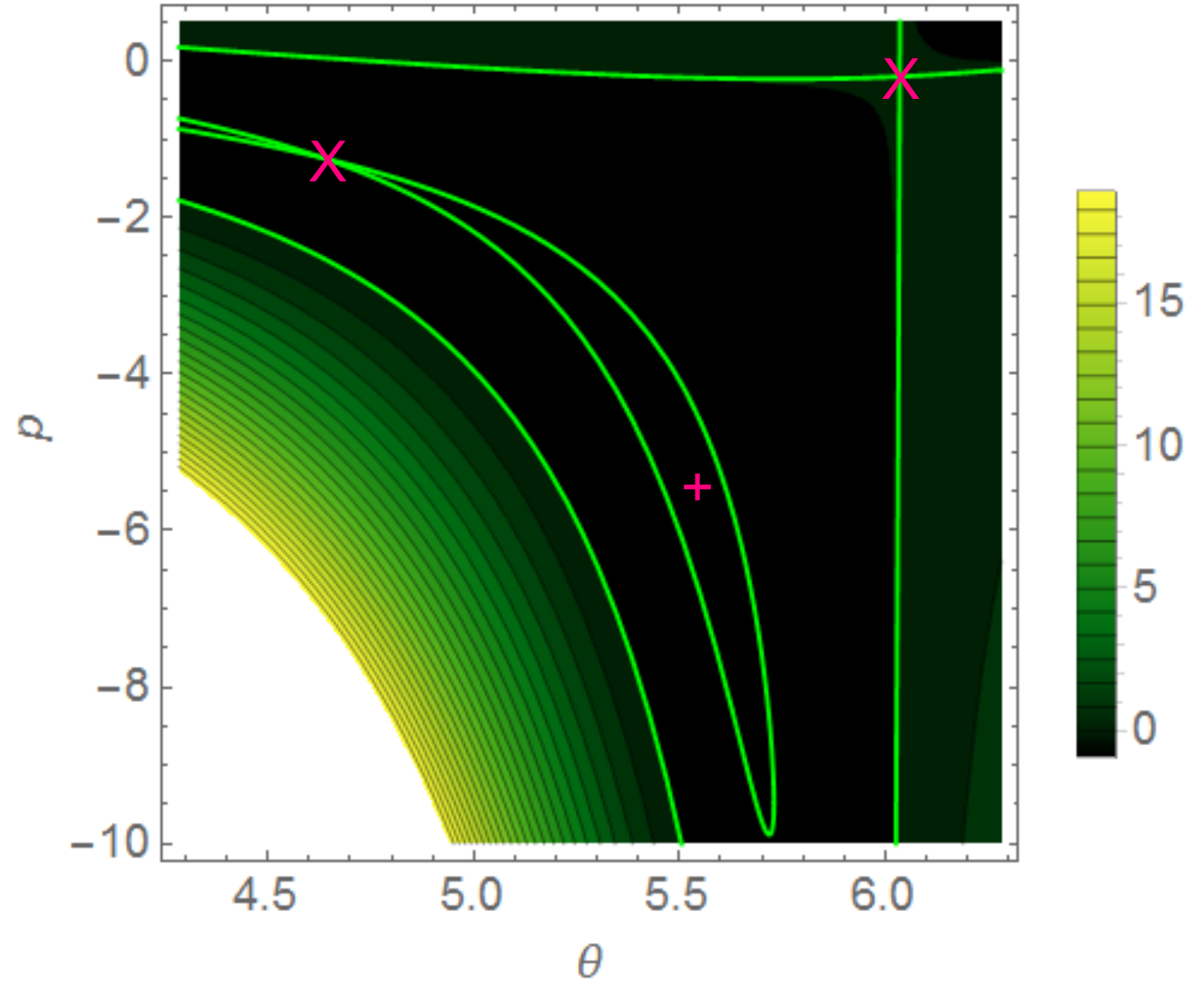}
\end{tabular} \begin{picture}(1,1)
\put(210,39){\footnotesize $E$ (MHz)}
\put(210,231){\footnotesize $E$ (MHz)}
\end{picture} \vspace{-10pt}
\caption{ We include the phase portrait for $\omega = 0.13$ and $\gamma=1$ (in the regime with three fixed points), with a complete picture (top) and reduced range to zoom in on the features of greatest interest (bottom). Two of the fixed points are hyperbolic (pink $\times$), at $(\bar{\theta} = 6.04, \bar{p} = -0.201)$ and $(\bar{\theta} = 4.63, \bar{p}= -1.25)$, and one is elliptical (pink $+$), at $(\bar{\theta} = 5.55, \bar{p} = -5.63)$. The latter two points are responsible for partitioning off the new regions D and E inside region B. Periodic paths are possible within the island D, which is bounded by a separatrix of energy $E_s = -0.938$ MHz. We note that for smaller $\omega$ compared with Fig.~\ref{fig-ps1sd1}, we have a larger region B which now encroaches asymmetrically into A (making winding-number MMLPs less likely), and the region C has grown (indicating that it is easier for trajectories to travel against the weaker Rabi drive).} \label{fig-smallom}
\end{figure}

In Figs.\ \ref{fig3} and \ref{fig4} of the main text, we show an example of a winding number path from the four-dimensional phase space which moves from $(x_i,z_i) =(0, -0.97)$ to $(x_f,z_f) = (-0.62,0.21)$. By simply taking the angles involved, we can find a pure-state analog of this MMLP, with $\theta_i = \pi$ and $\theta_f^{(B)} = -1.24$ or $\theta_f^{(A)} = \theta_f^{(B)}-2\pi$ (such that the number of winding counts also match the case shown from the experimental data). Mapping the mixed-state coordinates to the surface of the Bloch sphere gives us $(x_i,z_i) = (0,-1)$ and $(x_f,z_f) = (-0.95,0.32)$. The pure-state analog reinforces the winding-number interpretation of the paths in the larger space.
\par In the pure state case, the faster of the two paths lives in region A of the phase space (as labeled in Fig.~\ref{fig-ps1sd1}). The slower of the two paths must live in region B, because all paths in region A between the single (but periodically-repeating) fixed point actually move through that region too fast (all are past the desired point $\theta_f^{(B)}$ by the time $T = 1.4\ \mu s$). We find the energy in region A which traverses the distance $\theta_i \rightarrow \theta_f^{(A)}$ in $T = 1.4\ \mu s$ to be $E_A = 11.09$ MHz. We find that $E_B = -4.09$ MHz meets the boundary conditions for $\theta_{f}^{(B)}$. 
% predicted path numbers from pure state are a bad approximation; our analog isn't good enough for that. We could add comments or not. $P_B / P_A = e^{S_B- S_A} \approx 0.1$ as opposed to $ \approx 1.4$ in experiment
%The same ratio of actions in the four-dimensional phase-space was $1.44$, which matched nicely with the experimentally observed breakdown of paths $366/256 = 1.42$ (where more trajectories grouped around the path with \emph{fewer} winding counts).

\par We make an additional observation that for initial states immediately to the left of the system's fixed point, as in the case above, we see a narrow caustic in the Lagrange manifold across region B. These paths will diverge into regions of very negative $\dot{S}$ within the caustic, meaning that the window of opportunity to actually observe MMLPs here will be somewhat short-lived, in addition to being restricted to a narrow range of initial and final states.

\section{Appendix: Caustics for slow drive}
 Suppose we take $\omega = 0.13$, such that we have three fixed points in our MLP phase space instead of just one. The phase-portrait for this case is shown in Fig.~\ref{fig-smallom}. As $\omega \rightarrow \omega_c$ from above, region B from Fig.~\ref{fig-ps1sd1} deforms towards the shape shown in Fig.~\ref{fig-smallom}. When $\omega$ passes below $\omega_c$, we see the creation of the two new fixed points which form an additional separatrix, bounding off regions D (a periodic island) and E, as labeled in Fig.~\ref{fig-smallom}, inside region B. The creation of an elliptic fixed point and surrounding island D of periodic MLPs is of particular interest with regards to MMLPs, because it forces the possibility of true caustics (bounded by a diverging Van-Vleck determinant), with relatively long-lived MMLPs with an onset time that could be predicted by the theory. This is true provided that the periods inside the island D are not uniform (\emph{i.e.} vary through the island as a function of energy). This will necessarily be the case, as paths near the edge will have to slow down near the fixed point in the separatrix. Then catastrophes will be forced to appear in the Lagrangian manifold as in the example in \cite{phil2016}. This island will form different shapes in the manifold from the one we have studied previously, however, because it only has a fixed point in one end, rather than two, and is asymmetrically shaped. Catastrophes should also be possible to a much great extent in region B, where it is possible to choose initial states such that the manifold wraps around the island.

\section{Appendix: Experimental Setup}
The transmon circuit ($E_J/ h=24.6$ GHz, $E_C/ h=270$ MHz) was fabricated using double angle evaporation of aluminum on a high resistivity silicon substrate. The circuit was placed at the center of a 3D aluminum waveguide cavity (dimensions $34.15 \times 27.9 \times 5.25$ mm$^3$) which was machined from 6061 aluminum.

The cavity geometry was chosen to be resonant with the lowest energy transition of the transmon circuit. The resonant interaction between the circuit and the cavity (characterized by coupling rate $g/2 \pi = 136$ MHz) results in hybrid states, as described by the Jaynes-Cummings Hamiltonian. The lowest energy transition of hybrid states ($\omega_q/2\pi=6.3$ GHz) can therefore be considered a ``one-dimensional" artificial atom because the radiative decay of the system is dominated by the cavity's coupling to a  50 $\Omega$ transmission line.  This radiative decay was characterized by the rate $\gamma=1.42$ $\mu$s$^{-1}$. Resonance fluorescence  from the ``artificial atom" is amplified by a near-quantum-limited Josephson parametric amplifier, consisting of a $1.5$ pF capacitor, shunted by a Superconducting Quantum Interference Device (SQUID) composed of two $I_0 = 1\ \mu $A Josephson junctions.
 The amplifier is operated with negligible flux threading the SQUID loop and produces 20 dB of gain with an instantaneous 3-dB-bandwidth of 20 MHz. The quantum efficiency was measured to be 45\% \cite{nagh16}. We drive the qubit by sending a resonant coherent signal via a weakly coupled transmission line characterized by a Rabi frequency of $\Omega/2\pi = 0.9$ MHz.
 
The initial state fidelity was limited by a 3\% thermal population of the excited state.  The readout fidelity was enhanced by transferring the excited state population to a higher excited state of the system before applying the readout pulse \cite{nagh16}. All tomography results were corrected for the readout fidelity of 80\%.

\section{Appendix: Statistical information}
In the main text, the finite number of post-selected trajectories contributes a statistical uncertainty in the experimental MLP.  We present this as a uncertainty band indicating the standard deviation of the averaged 39 paths (Fig.\ 2), 18 paths (Fig.\ 4c), and 13 paths (Fig.\ 4d). Although the predicted experimental curves are in nice agreement with predicted theory curves,there are slight deviations which we attribute to limited ensemble of trajectories and finite size of post selection windows.  In Fig.\ \ref{subfig1}  the error bars indicate the uncertainty in the fit to the relative distribution (Eq.\ \ref{eq:rel}).  For this fit, the values for $\sigma_1$ and $\sigma_2$ are determined from the distributions $H_1(\mathcal{E})$ and $H_2(\mathcal{E})$ and the fit determines the mean value of $P_1(0)/P_2(0)$.  The error bars of Fig.\ \ref{subfig1} indicate the standard error of this mean based on the number of bins of $\mathcal{E}$. The number of bins is $(16,36,22,26)$ respectively left to right in Fig.\ \ref{subfig1}b).

%\bibliographystyle{unsrt}
%\bibliography{bibfile}

\end{document}